\documentclass[twocolumn]{aastex631}

\usepackage{amsmath}
\usepackage{graphicx}
\usepackage{hyperref}
\hypersetup{colorlinks=true,citecolor=blue,linkcolor=blue,urlcolor=blue}
\usepackage{natbib}
\usepackage{threeparttable}

\begin{document}

\title{Effect of time-varying X-ray emission from stellar flares on the ionization of protoplanetary disks}

\author[0000-0002-0169-3211]{Haruka Washinoue}
\affiliation{Department of Earth and Space Science, Osaka University, 1-1 Machikaneyama, Toyonaka, Osaka, 560-0043, Japan}

\author[0000-0003-3882-3945]{Shinsuke Takasao}
\affiliation{Department of Earth and Space Science, Osaka University, 1-1 Machikaneyama, Toyonaka, Osaka, 560-0043, Japan}

\author[0000-0002-2026-8157]{Kenji Furuya}
\affiliation{Department of Astronomy, Graduate School of Science, University of Tokyo, Tokyo 113-0033, Japan}

\begin{abstract}
X-rays have significant impacts on cold, weakly ionized protoplanetary disks by increasing the ionization rate and driving chemical reactions.
Stellar flares are explosions that emit intense X-rays and are the unique source of hard X-rays with an energy of $\gtrsim10$ keV in the protoplanetary disk systems.
Hard X-rays should be carefully taken into account in models as they can reach the disk midplane as a result of scattering in the disk atmospheres. However, previous models are insufficient to predict the hard X-ray spectra because of simplifications in flare models.
We develop a model of X-ray spectra of stellar flares based on observations and flare theories. The flare temperature and nonthermal electron emissions are modeled as functions of flare energy, which allows us to better predict the hard X-ray photon flux than before.
Using our X-ray model, we conduct radiative transfer calculations to investigate the impact of flare hard X-rays on disk ionization, with a particular focus on the protoplanetary disk around a T Tauri star.
We demonstrate that for a flare with an energy of $ 10^{35}$ erg, X-ray photons with $\gtrsim 5$ keV increase the ionization rates more than galactic cosmic rays down to $z \approx 0.1R$.
The contribution of flare X-rays to the ionization at the midplane depends on the disk parameters such as disk mass and dust settling.
We also find that the 10-year-averaged X-rays from multiple flares could certainly contribute to the ionization.
These results emphasize the importance of stellar flares on the disk evolution.
\end{abstract}



\section{Introduction} \label{sec:intro}

The stellar ultraviolet (UV) and X-ray radiations drive photoevaporation and photoionization, impacting on the disk evolution \citep{Nomura2007, Walsh2012, Ercolano2013, Nakatani2022}.
The disk accretion structure depends on the ionization rate via non-ideal magnetohydrodynamic effects \citep{Wardle2007, Bai2017, Suriano2019, Iwasaki2024}, and the disk chemical structure is also sensitive to it \citep{Aikawa1999, Ilgner2006a, Notsu2021}. 
Therefore, revealing the properties of the stellar ionizing photons and their impacts on the disks is of great importance. Recently, monitoring observations of disks have been performed for this purpose \citep[e.g.,][]{Cleeves2017, Terada2023}.

The X-ray activities in pre-main-sequence stars have been extensively studied by \textit{Chandra} and \textit{XMM-Newton} observations \citep{Imanishi2003, Getman2005, Feigelson2007, Gudel2007, Pillitteri2010}.
The X-rays in T Tauri stars (TTSs) are produced in magnetized coronae \citep[stellar coronal heating,][]{Gudel2004}, accretion shocks \citep[e.g.,][]{Gunther2007}, and stellar flares. The first two components may account for the quasi-steady component with a temperature less than a few keV. The temperature of magnetically heated coronae could be limited by the virial temperature \citep[e.g.,][]{Takasao2020} or the rate of the continuous energy injection from the stellar surface \citep[e.g.,][]{Shoda2021, Washinoue2023}. 
The stellar flares are explosions powered by magnetic reconnection and generate very hot ($\gtrsim10^{7-8}$ K) plasmas \citep{Shibata2011, Priest2014}. Therefore, in the protoplanetary disk systems, stellar flares should be the unique source of hard X-rays with an energy of $\gtrsim10$~keV. The energy of TTS flares is estimated to be $10^{33}-10^{37}$ erg \citep{Wolk2005, Lin2023}. Considering the maximum solar flare energy in history \citep[$\sim 10^{32}$ erg,][]{Shibata2013}, TTSs produce much stronger flares than the Sun. The occurrence rate of flares is also much higher in TTSs than in the Sun \citep{Wolk2005, Getman2021, Lin2023}, probably because of their rapid stellar rotation \citep{Gudel2004}. For stellar flares with an energy of $10^{35}$ erg, the peak X-ray luminosity can reach $\sim 10^{31}$ erg s$^{-1}$, which corresponds to approximately $10^{-2}$ of the solar bolometric luminosity.

Hard X-rays, together with cosmic rays, have impacts on the condition at and around the disk midplane.
For photons with an energy approximately above 10 keV, the scattering dominates the absorption in opacity \citep{Igea1999, Bethell2011}. As a result, a fraction of stellar hard X-rays scattered in the disk atmosphere can penetrate toward the midplane. This is not applied to soft X-rays and UV; they can only reach the disk surfaces due to absorption.

Time-variable hard X-rays from stellar flares will open a new window for the astrochemistry science. The chemical reaction rates need to be determined with certainty in models, but $\sim80$ \% of the total reactions has been still undetermined \citep{Tinacci2023, Balucani2024}, which is a big obstacle for comparison between theories and observations. Testing the theoretically predicted reaction rates is difficult in steady disks because of the lack of causality. However, chemical reactions driven by time-variable stellar hard X-rays could provide a great opportunity for the investigation.

Monitoring observations have shown some examples that time-variable stellar ionizing photons are affecting disks. \citet{Cleeves2017} reported an increase in the intensity of H$^{13}$CO$^+$ emission in response to enhanced X-ray luminosity in IM Lup system. Motivated by this observation, \cite{Waggoner2022} (hereafter referred to as WC22) studied the impact of flares with short and long timescales on the chemical changes in the disk. This is an extension of previous work by \cite{Ilgner2006}, which argued that the time-dependent X-ray luminosity could change the dead-zone structure.

There are theoretical studies that investigate the effect of time-variable stellar X-rays on the disks \citep{Ilgner2006, Waggoner2022}, but their flare models are insufficient to accurately predict hard X-ray photon spectra. The flare temperature controls the thermal component of the hard X-rays and is known to depend on flare energies \citep{Oka2015, Warmuth2016}. However, previous studies fix the flare temperature regardless of flare energies (e.g., WC22). In addition, solar flares, a prototype of stellar flares, commonly show nonthermal component of hard X-rays in the energy range of $\gtrsim 20$ keV \citep{Grigis2004, Oka2015}. The component is considered to be the bremsstrahlung emission by nonthermal electrons. Only considering the thermal emission from a flare temperature plasma significantly underestimates the flare hard X-rays because of a great contribution of the nonthermal emission. The nonthermal emissions have also been ignored in previous models.

This study aims to develop a better model of the X-ray spectra of stellar flares and to investigate the potential influence of hard X-rays on the disk ionization. 
For this purpose, we utilize the stellar flare theories and empirical relations based on previous flare observations. 
With our new X-ray model, we perform radiative transfer calculations of a protoplanetary disk to understand the role of stellar flares on the disk ionization condition.

This paper is organized as follows.
In Section \ref{sec:x_model}, we present the time-varying X-ray model for stellar flares to produce light curves and spectra.
We describe our disk model in Section \ref{sec:disk} and show the results of disk ionization due to X-ray flares in Section \ref{sec:result}.
We discuss our results and summarize the paper in Section \ref{sec:summary}.

\section{An X-ray model of stellar flares} \label{sec:x_model}

\subsection{X-ray light curves} \label{sec:lc}

X-ray light curves of stellar flares generally show a rapid rise followed by an exponential decay \citep[e.g.,][]{Getman2008, Aschwanden2012}.
If we express the light curve by a combination of the two exponential functions, the X-ray energy released by a flare can be written as 
\begin{align}
E_{\rm X} = L_{X_{\mathrm{peak}}} \left( \int_{-\infty}^{t_0} e^{t/\tau_{\rm rise}} dt + \int_{t_0}^{\infty} e^{-t/\tau_{\rm decay}} dt \right), 
\label{eq_ex}
\end{align}
where $L_{X_{\mathrm{peak}}}$ is the peak X-ray luminosity, and $\tau_{\rm rise}$ and $\tau_{\rm decay}$ are the rise and decay timescales of a flare, respectively.
Defining the fraction of the energy emitted in X-rays to the flare energy $E_{\rm flare}$ (the total energy released in a single flare) as $f_X$, we write $E_{\rm X}$ as
\begin{align}
E_X = f_X E_{\rm flare}.
\label{eq_fx}
\end{align}
With this simplified formulation, only $\tau_{\rm rise}$ and $\tau_{\rm decay}$ determine the shape of the light curves. Observations have suggested these two quantities are dependent on the flare energy $E_{\rm flare}$. We describe their energy dependencies based on observations in the following.

Observations of sun-like flares approximately indicate the following dependence of $\tau_{\rm rise}$ on $E_{\rm flare}$ \citep{Maehara2015}:
\begin{align}
\tau_{\rm rise} \propto E_{\rm flare}^{1/3}.
\label{eq_t_ris}
\end{align}
As discussed in \citet{Maehara2015}, this relation is consistent with the prediction of the reconnection flare model.
Regarding the coefficient of Relation~(\ref{eq_t_ris}), we utilize the results of the statistical survey on pre-main-sequence flares by \cite{Getman2021} by assuming that the TTS flares are similar to or scale-up versions of solar and sun-like stellar flares. Although we find a large scatter in data points, our fitting to their data suggests the following relation:
\begin{align}
\tau_{\rm rise} \approx 1.1\times 10^3 \left(\frac{E_{\rm flare}}{10^{35} \rm erg}\right)^{1/3} \rm s.
\label{eq_tris_g}
\end{align}

\cite{Getman2021} showed a correlation between $\tau_{\rm rise}$ and $\tau_{\rm decay}$. By combining this correlation and Relation~(\ref{eq_tris_g}), we obtain the relation for $\tau_{\rm decay}$:
\begin{align}
\tau_{\rm decay} \approx 7.5\times 10^3 \left(\frac{\tau_{\rm rise}}{1.1\times 10^3 \rm s}\right)^{0.44} \rm s.
\label{eq_tdec}
\end{align}
Again, note that this relation is based on data with a large scatter.
From Relations (\ref{eq_tris_g}) and (\ref{eq_tdec}), we derive $\tau_{\rm decay} \propto E_{\rm flare}^{0.15}$, yielding $L_{X_{\rm peak}} \sim E_{\rm flare} / \tau_{\rm decay} \propto E_{\rm flare}^{0.85}$.

\begin{figure}[t]
    \centering
    \includegraphics[width=8cm]{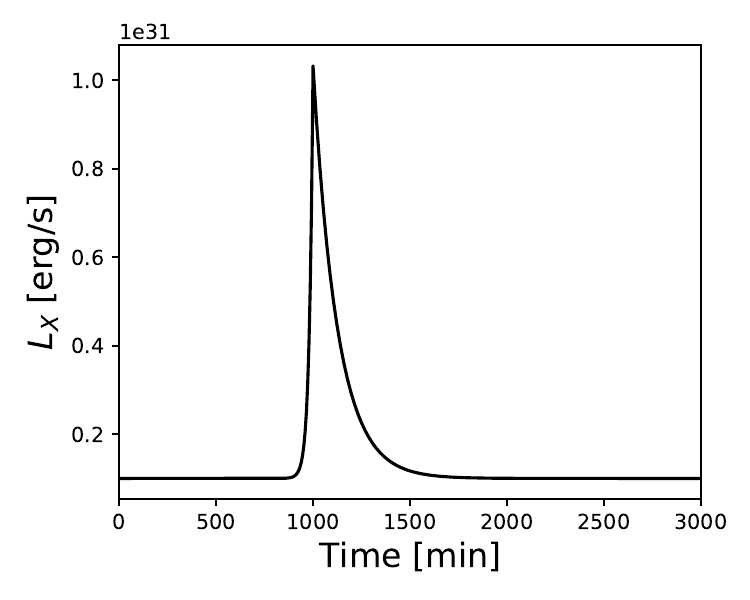}
    \caption{The X-ray light curve for $E_{\rm flare}=10^{35}$ erg.}
\label{fig:lc_1e35s}
\end{figure}
Now, we are ready to construct an X-ray light curve as a function of time $t$ and flare energy $E_{\rm flare}$. Let us define $L_{\rm ch}$ as the characteristic X-ray luminosity of TTSs in non-flaring phases (i.e. the steady component). $L_{\rm ch} = 10^{30}$ erg s$^{-1}$ is adopted in this study \citep{Flaccomio2003}. The light curve of a stellar flare which occurs at $t=0$ can be expressed as
\begin{align}
L_X(t) = L_{\rm ch} + L_{X_{\mathrm{peak}}} \times  \left\{\begin{array}{ll} \exp(t/\tau_{\rm rise}) & (t<0) \\ \exp(-t/\tau_{\rm decay})& (t\geq0). \end{array}
  \right.
\label{eq_lxt}
\end{align}
Figure \ref{fig:lc_1e35s} displays an example for the case with $E_{\rm flare}=10^{35}$~erg and $f_X=0.5$.
We have confirmed that our light curve model is generally consistent with typical solar flares. 
For instance, a GOES X1 class\footnote{The magnitudes of solar flares are classified to the GOES X-ray class (in ascending order of energy, B, C, M and X classes) according to the soft X-ray flux in the 1-8 \AA  band.} solar flare is defined as a flare whose peak X-ray luminosity is $3 \times 10^{26}$~erg~s$^{-1}$.
The typical bolometric energy of an X1-class flare is estimated to be of the order of $10^{31}$~erg. The flare duration (the sum of the rise and decay timescales) is of the order of hours. Our model can successfully produce a light curve consistent with a typical X1-class solar flare for the corresponding energy.

The X-ray energy fraction $f_X$ is a free parameter in our model.
It has been poorly constrained with the statistical studies even for solar flares \citep[e.g.,][]{Emslie2012, Aschwanden2017, Warmuth2020}.
We adopt $f_X=0.5$ in this work, which is derived to best fit Relation (\ref{eq_t_ris}) for $E_X$ and $\tau_{\rm decay}$ obtained by \cite{Getman2021}.


\subsection{X-ray spectra} \label{sec:xspec}

In addition to the X-ray luminosity, the spectral shape is also critical for the hard X-ray photon flux. This subsection describes how we construct the model of the X-ray spectrum.
The X-ray spectra of solar flares can commonly be fitted by a combination of single-temperature thermal components and nonthermal components produced by nonthermal electrons \citep{Emslie2004, Grigis2004, Warmuth2009, Oka2015}, although there are some arguments that the fitting based on the kappa distribution seems to be more physically reasonable \citep{Oka2018}. 
Considering the observational characteristics, we construct a flare spectrum that includes the two components under observational and theoretical constraints.

We define the total X-ray photon flux as a combination of the thermal and nonthermal radiation fluxes ($F_{\rm th}$ and $F_{\rm nth}$, respectively) as follows;
\begin{align}
  F_{\rm tot} &= \alpha F_{\rm th}(E_{\rm ph}) + \beta F_{\rm nth}(E_{\rm ph}),
  \label{eq_ftot}
\end{align}
where $E_{\rm ph}$ is a photon energy. 
$\alpha$ and $\beta$ are the nondimensional coefficients that determine the energy ratio of the nonthermal to thermal fluxes, $f_{\rm nth}$ (defined later). 
These two parameters are related by the following equation;
\begin{align}
  \frac{L_X}{4\pi r_0^2} = \alpha \int_{E_0}^{\infty}F_{\rm th}(E_{\rm ph})dE_{\rm ph} + \beta \int_{E_{\rm lc}}^{E_1} F_{\rm nth}(E_{\rm ph})dE_{\rm ph} .
  \label{eq_flux}
\end{align}

Defining the ratio of nonthermal flux to the thermal flux as 
\begin{align}
  f_{\rm nth} = \frac{\beta \int_{E_{\rm lc}}^{E_1} F_{\rm nth}(E_{\rm ph})dE_{\rm ph}}{\alpha \int_{E_0}^{\infty}F_{\rm th}(E_{\rm ph})dE_{\rm ph}},
  \label{eq_frac}
\end{align}
$\alpha$ and $\beta$ are calculated as follows;
\begin{align}
  \alpha = \frac{L_X}{4\pi r_0^2} \frac{1}{(1+f_{\rm nth}) \int_{E_0}^{\infty}F_{\rm th}(E_{\rm ph})dE_{\rm ph}},
  \label{eq_alpha}
\end{align}
\begin{align}
  \beta = \frac{L_X}{4\pi r_0^2} \frac{1}{(1+f_{\rm nth}^{-1}) \int_{E_{\rm lc}}^{E_1} F_{\rm nth}(E_{\rm ph})dE_{\rm ph}}.
  \label{eq_beta}
\end{align}

In modeling X-ray spectra, we compute $\alpha$ and $\beta$ for a given $f_{\rm nth}$.
The value of $f_{\rm nth}$ is highly uncertain.
Depending on events or methods for estimate, it widely ranges from $f_{\rm nth} \sim \mathcal{O}(0.1)$ to $\mathcal{O}(10)$ \citep[e.g.,][and references therein]{Warmuth2020}.
We adopt $f_{\rm nth}=0.3$ for the fiducial value which is within the observed range.
Despite the various possibilities for $f_{\rm nth}$, we assume that the nonthermal components are minor compared to the thermal ones, for a conservative estimation.

\begin{figure}[t]
    \centering
    \includegraphics[width=8cm]{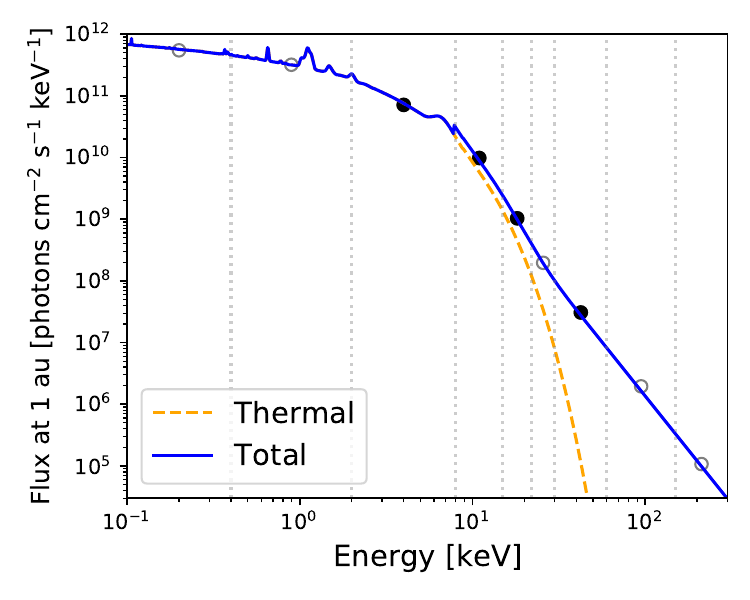}
    \caption{The X-ray spectrum at the peak of the flare with $E_{\rm flare}=10^{35}$ erg. The orange dashed and blue solid lines show the thermal and total (=thermal + nonthermal components) spectrum, respectively. The circles correspond to the mean flux of each energy bin (Equation (\ref{eq:meanf})).  }\label{fig:xspec_1e35}
\end{figure}

The thermal photon flux, $F_{\rm th}$, depends on the flare temperature and emission measure, as we assume that the thermal component shows a single temperature only. 
To calculate the flare temperature, we follow the method of \cite{Yokoyama1998}. 
They derived a scaling relation of the flare temperature $T_{\rm flare}$ by assuming the energy balance in reconnected magnetic fields. 
Namely, the reconnection heating and thermal conduction cooling by electrons are balanced in a reconnected loop. 
The derived relation is as follows:
\begin{align}
  T_{\rm flare} &\approx 3\times 10^7 \nonumber\\
  &\times \left(\frac{B}{50 \rm G}\right)^{6/7} \left(\frac{n}{10^9 \mathrm{cm}^{-3}}\right)^{-1/7} \left(\frac{L}{10^9 \rm cm}\right)^{2/7} \hspace{3mm} \mathrm{K} .
  \label{eq_temp}
\end{align}
If we can assume that the density and magnetic field strength in the preflare coronae do not vary significantly for different flares, only the flare size $L$ is the parameter that determines the flare temperature. 
We can relate the flare size to the flare energy as follows:
\begin{align}
  E_{\rm flare} \approx \frac{B^2}{8\pi} L^3 .
  \label{eq_emax}
\end{align}
Solar and sun-like stellar observations have shown that this relation is a good approximation of the upper end of the flare energy to the stellar spot area relation \citep[e.g.][]{Shibata2013}.
Using Equation~(\ref{eq_emax}), we rewrite $T_{\rm flare}$ as a function of $E_{\rm flare}$:
\begin{align}
  T_{\rm flare} &\approx 3.5 \times 10^7 \nonumber \\
  &\times  \left(\frac{n}{10^9\mathrm{cm}^{-3}}\right)^{-1/7} \left(\frac{B}{50 \rm G}\right)^{2/3} \left(\frac{E_{\mathrm{flare}}}{10^{35} \rm erg}\right)^{2/21} \hspace{3mm} \mathrm{K}.
  \label{eq_temp2}
\end{align}
We adopt typical values of solar flares for $n=10^9$ cm$^{-3}$ and $B=50$ G and assume that they are constants.

\begin{table}[t]
\scalebox{0.8}{ 
\begin{threeparttable}
\caption{The fiducial parameters for the X-ray light curve and spectrum.}
  \label{table:spec_param}
\begin{tabular}{lcc}
\hline
 Parameter & Symbol & Value \\
\hline \hline
 Fraction of X-ray energy in total\tnote{a} & $f_X$ & 0.5 \\
 Fraction of nonthermal to thermal energy\tnote{b}& $f_{\rm nth}$ & 0.3 \\
 Minimum energy\tnote{c} & $E_0$  & 0.1 keV \\
 Maximum energy\tnote{c} & $E_1$ & 300 keV \\
 Index of power-law relation (Equation (\ref{eq_fnth})) & $\gamma$ & 3.5 \\
\hline
\end{tabular}
\begin{tablenotes}
\item[a] Defined in Equation (\ref{eq_fx}).
\item[b] Defined in Equation (\ref{eq_frac}).
\item[c] Defined in Equation (\ref{eq_ftot}).
\end{tablenotes}
\end{threeparttable}
}
\end{table}

\begin{table*}[t]
\begin{threeparttable}
\caption{Comparison of the X-ray models between this study and \cite{Waggoner2022}.}
  \label{table:comp}
\begin{tabular}{lcc}
\hline
  & Waggoner \& Cleeves (2022) & \textbf{This work} \\
\hline \hline
 $\tau_{\rm rise}$, $\tau_{\rm decay}$ & fixed (3, 8 hours) & dependent on $E_{\rm flare}$\tnote{a}  \\
 Flare temperature & fixed  & dependent on $E_{\rm flare}$\tnote{b}  \\
 X-ray spectrum & observed IM Lup spectrum (1-20 keV)  & modeled with hard X-ray photons (0.1-300 keV) \\
\hline
\end{tabular}
\begin{tablenotes}
\item[a] Equations (\ref{eq_tris_g}) \& (\ref{eq_tdec}).
\item[b] Equation (\ref{eq_temp2}).
\end{tablenotes}
\end{threeparttable}
\end{table*}
\begin{figure*}[t]
    \centering
    \includegraphics[width=16cm]{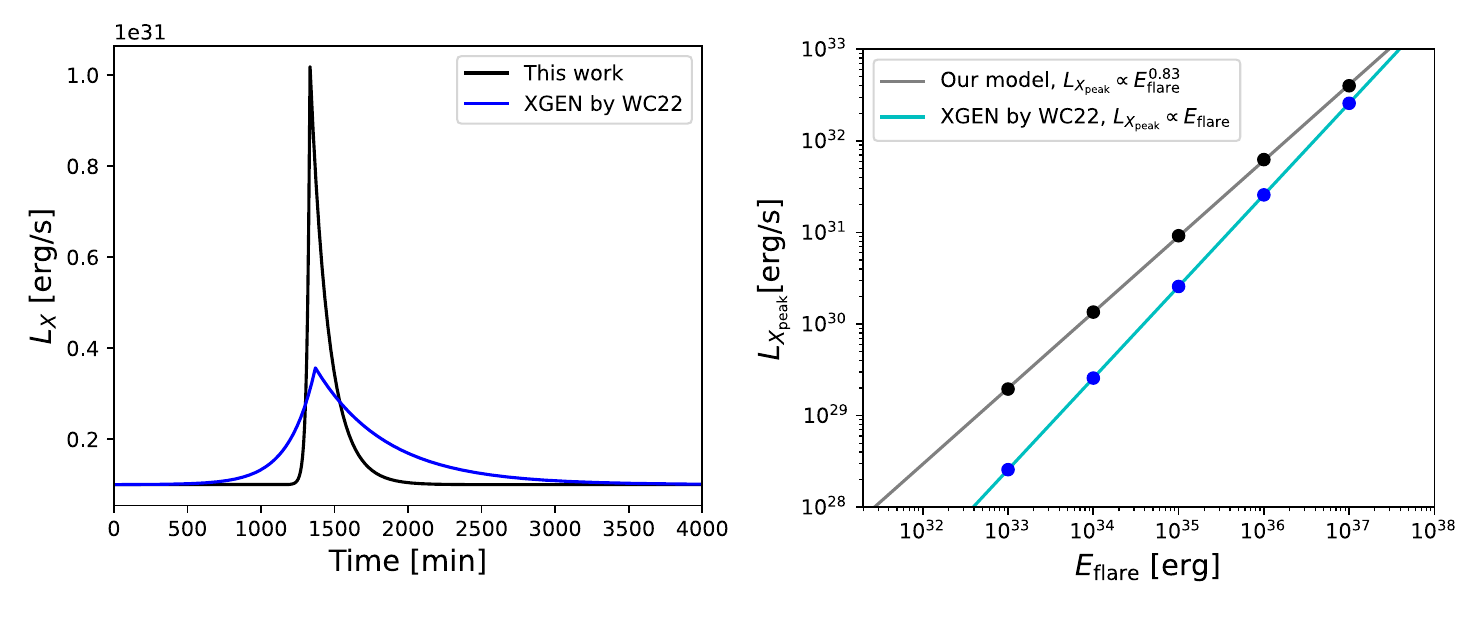}
    \caption{Left: The X-ray light curves for $E_{\rm flare}=10^{35}$ erg. The different models are compared with black (this work) and  blue (XGEN by WC22) lines. Right: $L_{X_{\rm peak}}$ vs. $E_{\rm flare}$ for the two models.}\label{fig:lc_1e35}
\end{figure*}

The emission measure is approximately expressed as 
\begin{align}
  \mathrm{EM} \approx n^2 L^3.
  \label{eq_em0}
\end{align}
By combining Equations (\ref{eq_temp}) and (\ref{eq_em0}), we can rewrite EM as
\begin{align}
\begin{split}
  \mathrm{EM} &\approx 4.2\times 10^{52} \\
 &\times \left(\frac{B}{50\mathrm{G}}\right)^{-5} \left(\frac{n}{10^9\mathrm{cm}^{-3}}\right)^{3/2} \left(\frac{T_{\rm flare}}{3.5\times10^7 \mathrm{K}}\right)^{17/2} \mathrm{cm}^{-3}
  \label{eq_em}
\end{split}
\end{align}
\citep{Shibata1999, Shibata2002}.
The validity of Equations (\ref{eq_temp}) \& (\ref{eq_em}) have been supported by observations \citep{Namekata2017, Aschwanden2020} and numerical simulations \citep{Yokoyama2001, Shiota2005, Takasao2015}.
With these quantities, we compute $F_{\rm th}$ with Chiantipy v0.15.1 from CHIANTI database version 10.0 \citep{DelZanna2021}.
The CHIANTI package contains a large set of atomic data and has been widely used to compute optically thin synthetic spectra with the continuum and line emissions from astrophysical sources \citep[e.g.,][]{Shoda2021}.

In producing the nonthermal spectrum, we utilize the empirical equation from observations of solar flares, as there are no established theoretical models \citep[e.g., see a review of][]{Oka2018}.
Observed spectra from solar flares show the misalignment with thermal component at $\gtrsim20$ keV.
The deviated component is interpreted as a nonthermal emission because it can be well fitted with a single power-law function in many cases.
Previous observations have reported a wide variability in the index of the power-law relation and the ratio of nonthermal component in total emission.
The power-law index typically resides in $2\lesssim \gamma\lesssim 7$ \citep[][and references therein]{Oka2018}.

We adopt the single power-law formulation for the nonthermal X-ray flux described as follows \citep[e.g.,][]{Grigis2004}:
\begin{align}
F_{\rm nth}(E_{\rm ph}) = F_{E_0} \left(\frac{E_{\rm ph}}{E_0}\right)^{-\gamma} \hspace{3mm} \mathrm{at} \hspace{3mm} E_{\rm ph}>E_{\rm lc},
\label{eq_fnth}
\end{align}
where $E_{\rm ph}$ indicates the photon energy. 
$E_{\rm lc}$ represents a low-energy cutoff, and $F_{E_0}$ indicates the normalization flux value at $E_0$. We adopt $E_{\rm lc}\sim10$~keV, $F_{E_0}=0.525$~photons cm$^{-2}$ s$^{-1}$ keV$^{-1}$ and $E_0=50$~keV as fiducial values.
We set $\gamma=3.5$ as a fiducial index based on the observational surveys of solar flares \citep{Oka2013, Oka2018}, despite the fact that the large variation is found as mentioned above.

Figure \ref{fig:xspec_1e35} shows the X-ray spectrum at the peak of the flare with $E_{\rm flare}= 10^{35}$~erg.
The parameters used for this model are summarized in Table \ref{table:spec_param}.
The orange-dashed and blue-solid lines denote the thermal and total fluxes, respectively.
The comparison between the two lines indicates that the nonthermal component dominates in the total X-ray flux in the range of $E_{\rm ph}\gtrsim 10$~keV.
With the nonthermal component, the photon flux at $E_{\rm ph}\gtrsim 10$~keV is a few orders of magnitude larger than that in the case only with the thermal component. 
We note that a small bump at $E_{\rm ph}=E_{\rm lc}$ is an  artificial structure. We will be able to minimize the size of the bump by introducing a smoothing function for the nonthermal spectrum around the cut-off energy.

\subsection{Comparison with the model by Waggoner \& Cleeves 2022}

We compare our model with the X-ray flare model by WC22.
The key contrasts are summarized in Table \ref{table:comp}.
WC22 developed a X-ray lightcurve model, named XGEN.
A notable difference in generating light curves between the two models is the setting of $\tau_{\rm rise}$ and $\tau_{\rm decay}$.
In our model, they depend on $E_{\rm flare}$ according to Equations (\ref{eq_tris_g}) and (\ref{eq_tdec}), while XGEN adopts the fixed values of $\tau_{\rm rise}=3$ hours and $\tau_{\rm decay}=8$ hours.
This results in different $L_{X_{\rm peak}}$ between the models for a given $E_{\rm flare}$.
The left panel in Figure \ref{fig:lc_1e35} compares the light curves using our model (black line) and XGEN (blue line) in case of $E_{\rm flare}=10^{35}$ erg.
We see that $L_{X_{\rm peak}}$ for our model is approximately 3 times larger than that for XGEN in this case.
The dependencies of $L_{X_{\rm peak}}$ on $E_{\rm flare}$ for the two models are plotted in the right panel of Figure \ref{fig:lc_1e35}.
The discrepancy of $L_{X_{\rm peak}}$ in the two models is more significant for smaller $E_{\rm flare}$. 
Considering that the flare occurrence rate is higher for weaker flares (see Section \ref{sec:flares}), this discrepancy will result in a large difference in a long-term evolution.
As shown in Section~\ref{sec:lc}, our model predicts that $L_{X_{\rm peak}} \sim E_{\rm flare} / \tau_{\rm decay} \propto E_{\rm flare}^{0.85}$.
This is in a good agreement with the fitting function for our data, $L_{X_{\rm peak}} \propto E_{\rm flare}^{0.83}$ (black line).
On the other hand, XGEN shows a stronger dependency of $L_{X_{\rm peak}} \propto E_{\rm flare}$ (cyan line) because $\tau_{\rm decay}$ is independent on $E_{\rm flare}$.

\begin{figure}[t]
    \centering
    \includegraphics[width=8cm]{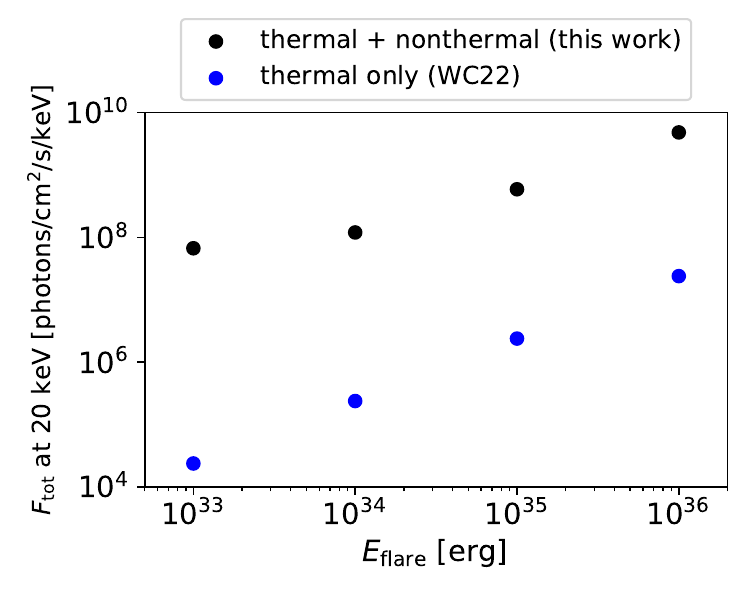}
    \caption{The photon flux values at 20 keV against $E_{\rm flare}$. The black dots show data of our model which incorporates both thermal and nonthermal components. The blue dots indicate data of the thermal-only case adopted in WC22.}\label{fig:fnth}
\end{figure}

The X-ray-spectrum modeling is also different between the two models.
WC22 adopt the observed IM Lup spectrum (1-20 keV) as a reference spectral shape, which can be fitted by a combination of three thermal components \citep{Cleeves2017}. They fix the spectral shape for flares with different flare energies, which means that the flare temperature is assumed to be independent of the flare energy. Although their model spectrum is based on an actual observation, the model is missing the general properties of stellar flares. On the other hand, we model the spectral shape based on the theories and observations of stellar flares (Section \ref{sec:xspec}).
For a better modelling of hard X-ray spectrum, we have improved the determination of the flare temperature based on the flare theories. The nonthermal component is also included by referring to solar and stellar observations. 

We compare the photon flux at hard X-ray regime between the two models.
Figure \ref{fig:fnth} shows the photon flux values at 20~keV for different $E_{\rm flare}$.
The black points show data of our model with thermal and nonthermal components, where $f_{\rm nth}=0.3$.
The blue dots are obtained by the spectrum adopted in WC22, where the magnitude of the spectrum is scaled with the flare energy uniformly with respect to the photon energy based on the reference spectrum.
Our model shows a much larger photon flux at 20~keV by two to three orders of magnitudes than WC22 model, which emphasizes the importance of the contribution of nonthermal electrons.
For a better modelling of hard X-ray spectrum, we have improved the determination of the flare temperature based on the flare theories. The nonthermal component, which has been ignored in previous studies, is also included by referring to solar and stellar observations. With these improvements, our model is capable of covering a wider range of the photon energy than the previous model, which is from 0.1 to a few 100~keV. This study presents the results based on the spectra in the range of 0.1 to 300~keV.

\section{Setup of radiative transfer calculations} \label{sec:disk}

To investigate how the X-ray radiation from stellar flares affects the disk ionization, we perform radiative transfer calculations with RADMC-3d code \citep{Dullemond2012} for a protoplanetary disk using our X-ray flare model (Section \ref{sec:xspec}).
The radiative transfer is solved in spherical coordinates $(r,\theta)$. We also use the cylindrical radius $R=r\sin\theta$ in the following.

\subsection{Star and disk models} \label{sec:d_model}

The star is approximated as a point source at the center of the system. The stellar radiation is isotropic, nevertheless the flare emission could be anisotropic. 
We adopt a parameter set for a typical classical TTS: The stellar mass $M_*=0.5$ $M_{\odot}$, radius $R_*=2.0$  $R_{\odot}$ and effective temperature $T_{\rm eff}=4000$ K.

The disk around the star is modeled as a Keplerian disk whose mass $M_{d}=0.01 M_{\odot}$.
The inner and outer radii of the disk are set to be 1.0 and 100 au, respectively.
The dust temperature at each grid cell is computed by thermal Monte Carlo radiative transfer calculation which assumes the local thermodynamic equilibrium for dust particles. 
The following equation determines the dust temperature:
\begin{align}
 \sigma_{\rm SB} T^4 = \frac{N L_*}{4N_{\gamma}\kappa_P m},
\end{align}
where $\sigma_{\rm SB}$ is the Stefan-Boltzmann constant, $N$ is the number of photon packets absorbed in the grid cell, $N_{\gamma}$ is the number of photon packets reaching the grid cell, $L_*$ is the stellar bolometric luminosity, $\kappa_P$ is the Planck mean opacity and $m$ is the dust mass in the grid cell \citep{Bjorkman2001}. 
Our model gives the midplane temperature of $T\sim 30$~K at $R=10$~au.
The density distribution is given from the hydrostatic condition.
We set the single dust size of $0.1 \hspace{1mm}\mu$m and assume a uniform dust-to-gas mass ratio of $0.01$.
We also perform the calculations with different models, including variations in disk mass, dust-settling effects, and the position of the X-ray source, to investigate their impact on disk ionization (see Appendix \ref{sec_app1} and \ref{sec_app2}.)

The grids are spaced logarithmically in the $r$ direction and linearly in the $\theta$ direction. The grid numbers are 256 in both directions. The minimum grid sizes in the $r$ and $\theta$ directions are $1.8\times 10^{-2}$ au and $2.73\times 10^{-3}$ in radians, respectively, which are sufficient to resolve the disk structure at all radii.

\subsection{Inputs for Radiative transfer calculations} \label{sec:rad}

This study only investigates the effect of stellar flare X-rays on the ionization degree. For this aim, we assume that the X-rays have no effect on the density and temperature structures of the disk and we perform the radiative transfer calculations for X-rays only.
We also assume that all the X-ray photons emitted from the central star directly reach to the disk without extinction from accretion flow or disk winds.
These approaches greatly simplify the calculations.

Our X-ray spectrum is a continuous function, but the spectrum is binned with respect to the photon energy so that we can express the flare X-ray spectrum as a combination of the multiple monochromatic spectra (Figure \ref{fig:xspec_1e35}). For each bin, we perform the monochromatic Monte Carlo calculation to obtain the mean intensity in the calculation domain. The total mean intensity for all the energy bins is calculated as the sum of the individual contributions. 


Each bin has its mean flux as
\begin{align}
\overline{F_i} = \frac{1}{\Delta{E_i}}  \int_{E_i - \Delta E_i /2}^{E_i + \Delta E_i /2} F_i(E_{\rm ph}) dE_{\rm ph}, 
\label{eq:meanf}
\end{align}
where $\Delta {E_i}$ and $E_i$ are the energy width and the mean energy of the bin $i$.
$\overline{F_i}$ is plotted with the circles in Figure \ref{fig:xspec_1e35}.
The values for $E_i - \Delta E_i /2$ are $0.1, 0.4, 2.0, 8.0, 15, 22, 30, 60$ and 150 keV for this figure (gray dotted lines).
The total mean X-ray luminosity is calculated by
\begin{align}
 \overline{L_X} = 4\pi d^2 \sum_i E_i \overline{F_i} \Delta E_i .
\end{align}
This binning method ensures that the photon flux in each bin is the same as the original spectrum.
The spectrum is split such that following relation is satisfied:
\begin{align}
 \frac{|L_X - \overline{L_X}|}{L_X} \leq 0.1 ,
 \label{eq_lx01}
\end{align}
where $L_X$ is the X-ray luminosity extracted from the modeled light curve (Section \ref{sec:lc}).
We confirm that Equation (\ref{eq_lx01}) holds even in the region where $E_i-\Delta E_i/2 \geq 20$ keV.
In the following part, we show the results for the 4 bins whose representative energy is 5.0, 10, 20 and 40 keV (black filled circles in Figure \ref{fig:xspec_1e35}).

\begin{figure}[t]
    \centering
    \includegraphics[width=8cm]{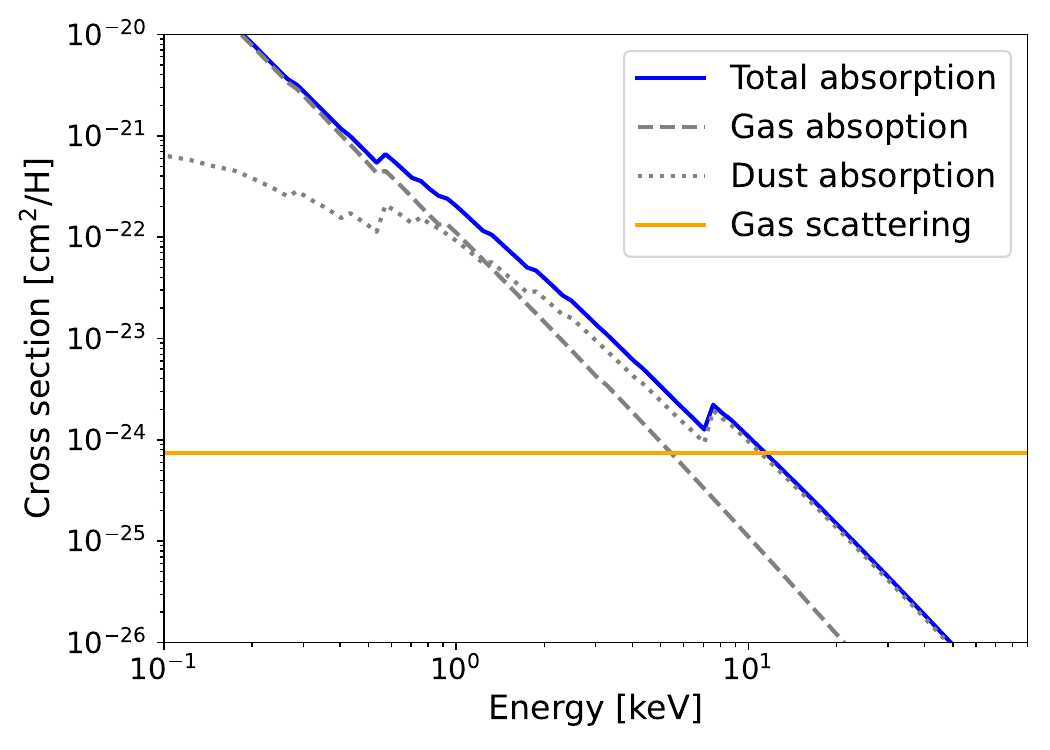}
    \caption{Absorption (blue) and scattering (orange) cross sections as a function of energy. The gray lines correspond to the absorption due to gas (dashed) and dust (dotted) components, whose sum yields the total absorption (blue). The scattering cross section is independent on energy (Thomson scattering). }\label{fig:cross_ab}
\end{figure}

The absorption ($\sigma_X(E_{\rm ph})$) and scattering opacities for mixtures of gas and dust are computed using xraylib library \citep{Brunetti2004} assuming the solar abundance and that each element is partitioned between gas and dust according to Table 1 of \cite{Bethell2011}.
These dependencies on energy are shown in Figure \ref{fig:cross_ab}.
The total absorption cross section (blue solid line) is the sum of gas (gray dashed line) and dust (gray dotted line) components. 
In this case, the scattering (orange solid line) is more efficient than absorption at energies above 10 keV.

\subsection{Calculation of ionization rates} \label{sec:xionization}

\begin{figure*}[t]
    \centering
    \includegraphics[width=16cm]{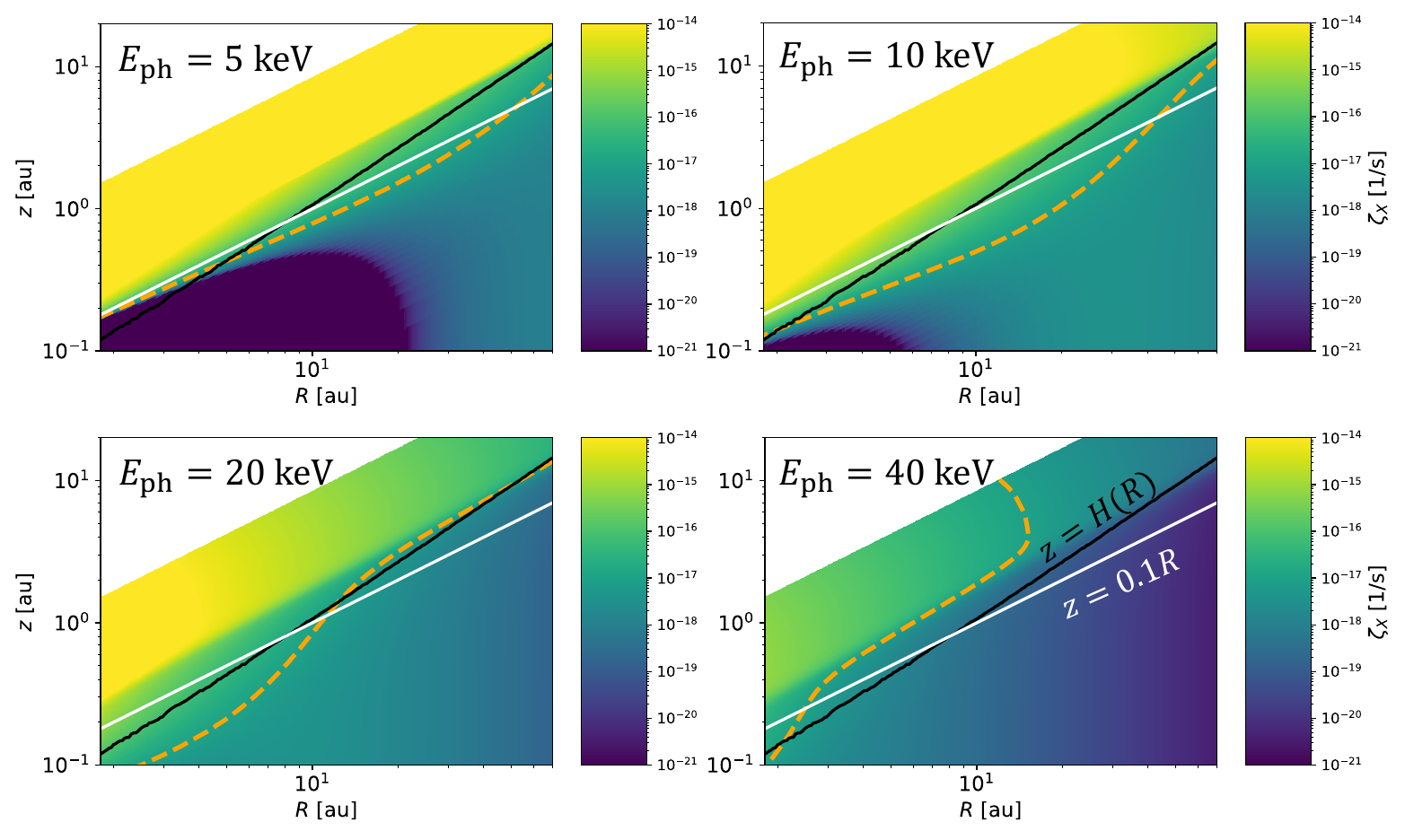}
    \caption{The ionization rate by photons with $E_{\rm ph}=5$ keV (upper left), 10 keV (upper right), 20 keV (lower left), and 40 keV (lower right) at the peak of flare with $E_{\rm flare}=10^{35}$ erg. The black lines represent the disk scale height described as $H(R) = \sqrt{2}c_s(R)/\Omega_K(R)$, where $c_s(R)$ is the isothermal sound speed and $\Omega_K (R)=\sqrt{GM/R^3}$ is the Keplarian angular velocity. The white solid line represents $z=0.1R$. The orange dashed line is the contour for $\zeta_X=10^{-17}$ s$^{-1}$.}\label{fig:irate_1e35}
\end{figure*}

The monochromatic Monte Carlo run for a given photon energy, $E_{\rm ph}$, gives the mean intensity as a function  in spherical coordinates:
\begin{align}
J_X = \frac{1}{4\pi} \int I_X d\Omega.
\label{eq_jx}
\end{align}
Using $J_X$, we can compute the X-ray ionization rate $\zeta_X$ as 
\begin{align}
\zeta_X = 4\pi \int \frac{\sigma_X(E_{\rm ph}) J_X}{\Delta \epsilon} dE_{\rm ph} \hspace{3mm} \mathrm{s}^{-1},
\label{eq_etax}
\end{align}
where $\sigma_X(E_{\rm ph})$ is the absorption cross section for X-ray photons (Figure \ref{fig:cross_ab}).
$\Delta \epsilon = 37$ eV is the mean energy required per secondary ionization \citep{Shull1985}.
Equation (\ref{eq_etax}) assumes that the Auger electrons with the ionization energy are produced by X-rays.
We calculate $\zeta_X$ for individual bins. The total ionization rate corresponds to the sum of all the contributions.

\section{Results of radiative transfer calculations}
\label{sec:result}
\subsection{Single flare case}
\subsubsection{The spatial structure of the ionization rate}

As a demonstration of our X-ray model, we calculate the ionization rates at the peak of the flare with $E_{\rm flare}=10^{35}$ erg.
The distribution of ionization rates for photon fluxes across different energy bins is shown in Figure \ref{fig:irate_1e35} (for the photon spectra, see Figure~\ref{fig:xspec_1e35}).
To highlight the relative importance of flare X-rays and galactic cosmic ray ionization, we plot the contour for $10^{-17}$ s$^{-1} (\equiv \zeta_{\rm CR,0})$ (dashed orange lines).
This value is in the range of typical values of the galactic cosmic ionization rate \citep[e.g.][]{Dalgarno2006}.

\begin{figure*}[t]
    \centering
    \includegraphics[width=16cm]{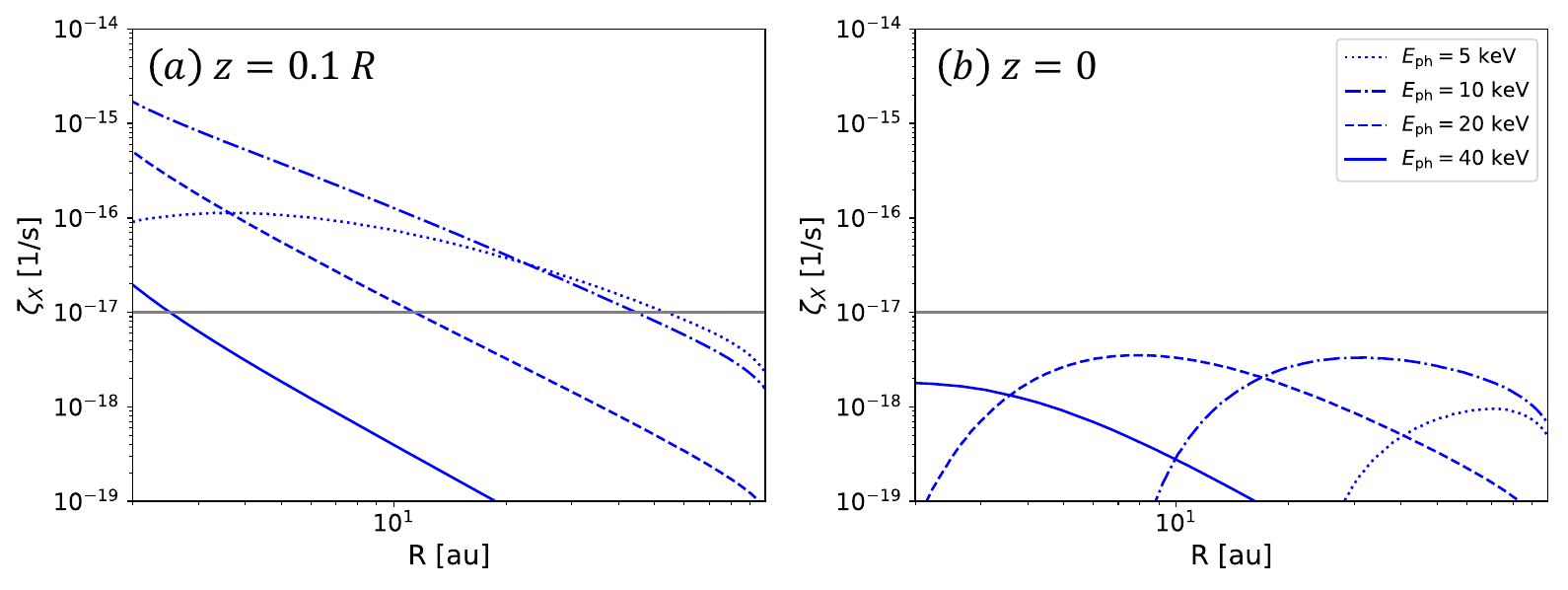}
    \caption{The ionization rates at (a) $z=0.1R$ and (b) $z=0$ at the peak of the flare with $E_{\rm flare}=10^{35}$ erg. The different line types represent different photon energy, $E_{\rm ph}$. The gray solid line is the typical value of cosmic ray ionization rate \citep{Dalgarno2006}.}\label{fig:jx_irate}
\end{figure*}

Figure~\ref{fig:irate_1e35} shows that X-rays with the energy of 10~keV enhance the ionization rate in the widest region of the disk (see also Figure~\ref{fig:jx_irate}). 
X-rays with lower energy (5~keV) have higher photon fluxes (Figure~\ref{fig:xspec_1e35}), but due to significant absorption, they cannot penetrate to $R\geq 20$ au. X-rays with the higher energies (20 and 40~keV) experience less absorption, and they can increase the ionization rate in the inner disk ($<$ a few au).
However, owing to their small photon fluxes, their contribution to the ionization rate is limited at outer radii ($>$ a few au). X-rays with the energy of 10~keV have modest photon fluxes and absorption cross-sections.

The radial profiles of $\zeta_X$ for different $E_{\rm ph}$ at $z=0.1R$ and $z=0$ (midplane) are shown in Figure~\ref{fig:jx_irate}.
Figure~\ref{fig:jx_irate}(a) displays that $\zeta_X$ exceeds $\zeta_{\rm CR,0}$ at nearly all radii, which suggests that the stellar X-ray flares can be a primary source of the ionization down to $z/R=0.1$. 
On the other hand, Figure~\ref{fig:jx_irate}(b) shows that $\zeta_X$ at $z=0$ is slightly lower than $\zeta_{\rm CR,0}$. This indicates a minor contribution of flare X-rays to disk ionization at the midplane in this case. However, the X-ray can increase the ionization degree at the disk midplane depending on the disk parameters. We present some results with different disk parameters in Appendix \ref{sec_app1}. 


\subsubsection{Dependence of flare energy}
\label{subsec:dep_ene}

We investigate regions of influence by stellar flares with different flare energies.
We define the critical radius for ionization as $r_{\rm crit}$, which satisfies

\begin{align}
\zeta_X(R=r_{\rm crit}, z) \geq \zeta_{\rm CR,0},
\label{eq:rcrit}
\end{align}
where we adopt $z= 0.1 R$ and $z=0$ in this study.
For the flare with $E_{\rm flare}=10^{35}$~erg, the critical radii denote the radii for the intersection points between blue and gray lines in Figure \ref{fig:jx_irate}. 
For instance, in Figure  \ref{fig:jx_irate}(a), X-ray photons with $E_{\rm ph}=5$~keV shows $r_{\rm crit}\approx 50$~au, while X-rays with $E_{\rm ph}=40$~keV indicates $r_{\rm crit}\approx 2.5$~au.
Note that it is possible that more than one critical radii can appear for a single photon energy. Namely, X-rays with each $E_{\rm ph}$ may have another critical radius inside $r\lesssim 1$~au, which is not covered by our disk model. 

Figure~\ref{fig:rcrit} displays the photon energy dependencies of $r_{\rm crit}$ for $z=0.1 R$ and $z=0$ for flares with different $E_{\rm flare}$. The upward triangles denote the critical radii above which $\zeta_X > \zeta_{\rm CR,0}$, while the downward triangles indicate the ones below which $\zeta_X > \zeta_{\rm CR,0}$. We call the critical radii denoted by upward and downward triangles as the inner and outer critical radii, respectively. 
In both cases with $z=0.1 R$ and $z=0$, the regions influenced by stellar flares extend when $E_{\rm flare}$ is large.
Figure~\ref{fig:rcrit}(a) shows the large outer critical radii for $E_{\rm ph}=5$ and 10~keV, as expected in Figures~\ref{fig:irate_1e35} and \ref{fig:jx_irate}.
The outer critical radii for $E_{\rm ph}\gtrsim20$~keV show a strong dependence on $E_{\rm flare}$. 
Stronger flares show larger photon fluxes at higher photon energies, which results in the increase of the outer critical radii.
Figure~\ref{fig:rcrit}(b) shows that X-rays from strong flares with $E_{\rm flare}=10^{36}$ and $10^{37}$ erg can significantly impact the ionization at the midplane. 
In this case, the X-rays with lower photon energies ($E_{\rm ph}\lesssim 20$~keV) have larger inner critical radii because the low energy X-ray photons are significantly attenuated by dusts and gas (Figure~\ref{fig:cross_ab}).


\subsection{Multiple flare case}
\label{sec:flares}

\begin{figure*}[!t]
    \centering
    \includegraphics[width=16cm]{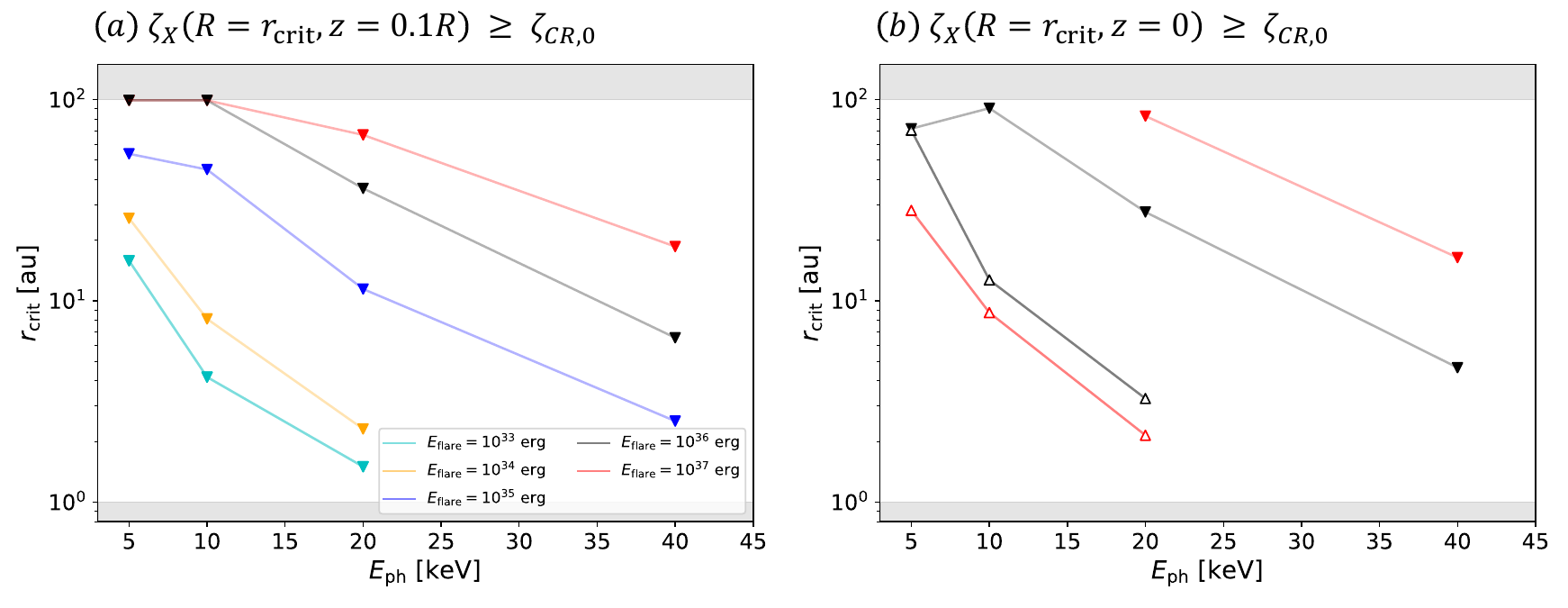}
    \caption{The photon energy dependence of $r_{\rm crit}$ for (a)$z=0.1R$ and (b)$z=0$. The data are taken at the flare peak. The different colors denote the results for flares with different $E_{\rm flare}$. The gray shaded regions indicates the regions which are not covered by our disk model. The upward triangles indicate the critical radii above which $\zeta_X > \zeta_{\rm CR,0}$, and vice versa.}\label{fig:rcrit}
\end{figure*}

Stellar flares are intrinsically intermittent, but their time-averaged property would be a key for the long-term evolution of protoplanetary disks. To study the time-averaged property, we model the flare occurrence frequency based on solar and stellar observations.


The solar observations have shown that the occurrence frequency of solar flares generally obeys the following power-law relation:
\begin{align}
\frac{dN}{dE_{\rm flare}} = A E_{\rm flare}^{-\alpha},
\label{eq:freq}
\end{align}
where $A$ is a normalization coefficient \citep{Datlowe1974, Dennis1985}.
Previous studies have typically found $\alpha\sim1.5-1.8$ in the range of $10^{24}$ erg $ <E_{\rm flare} <10^{32}$ erg \citep{Shimizu1995, Aschwanden2000, Jess2019}.

The stellar X-ray observations have implied similar power-law relations for the flare occurrence frequencies of solar-type main-sequence stars \citep[e.g.][]{Shibata2013, Shibayama2013}.
Similar results have also been obtained for pre-main-sequence stars \citep{Getman2021, Lin2023}.
We note that the flare occurrence frequencies for stellar flares are estimated not for a single star; flare data of multiple stars are combined to calculate the frequency because of observational limitations.
Furthermore, the frequency distribution towards the upper end of $E_{\rm flare}$ is expected to deviate from a power-law function \citep{Okamoto2021}.

Based on these observations, we assume that the flare frequency in the classical TTS follows Equation (\ref{eq:freq}) in the range of $10^{33}$ erg $\leq E_{\rm flare} \leq 10^{37}$ erg.
We adopt $\alpha = 1.8$, which is a typical value for solar flares \citep{Shibata2013}. 
We determine the normalization coefficient such that flares with $E_{\rm flare}=10^{34}$~erg occur $\sim 100$ times per year, which is in the range of observations of pre-main-sequence stars by \citet{Wolk2005}.

\begin{figure*}[t]
    \centering
    \includegraphics[width=17cm]{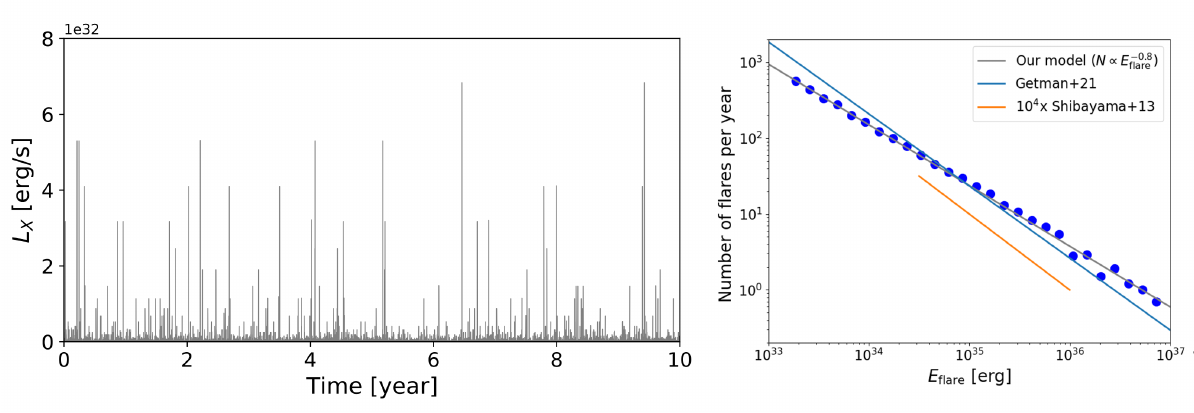}
    \caption{Left: The X-ray light curves over 10 years. Right: The number of flares per year. The blue dots are the data from our calculation and the gray solid line follows $N\propto E_{\rm flare}^{-0.8}$. The blue and orange lines correspond to the flare frequency distributions obtained by \cite{Getman2021} for pre-main-sequence stars and \cite{Shibayama2013} for slowly rotating G-type dwarfs ($P>10$ days), respectively. The flare number from \cite{Shibayama2013} is multiplied by $10^4$ for better visibility.}\label{fig:lc_10yr}
\end{figure*}

The left panel of Figure \ref{fig:lc_10yr} displays the X-ray light curve over the 10~years. 
The 10-year time-span is long enough to obtain the expected flare occurrence frequency in the range of $10^{33}~{\rm erg}< E_{\rm flare}< 10^{37}~{\rm erg}$.

The right panel shows the number of flares per year, which confirms that the data points nearly follow the prediction of the input power-law flare occurrence frequency ($N\propto E_{\rm flare}^{-0.8}$ from Equation (\ref{eq:freq}) with $\alpha=1.8$, denoted by the gray solid line).
To compare our flare-frequency model with flare observations, we plot the observed flare frequency for pre-main-sequence stars with the blue line \citep{Getman2021} and for slowly rotating G-type dwarfs ($P>10$ days) with orange line \citep[multiplied by $10^4$,][]{Shibayama2013}.


In addition to the spectrum from the single flare, the time-averaged spectrum from multiple flares is important to investigate the effective impact on the long-term disk evolution.
For this reason, we time-averaged the X-ray luminosity over the 10 years.
We note that the 10-year time-averaged is sufficient to understand the long-term disk evolution because the 10-year covers the full range of possible flare energy for pre-main-sequence stars, $10^{33}$ erg $\leq E_{\rm flare} \leq 10^{37}$ erg, whose frequency follows Equation (\ref{eq:freq}).
We expect that extending the time-span for averaging does not significantly alter the following results.
The resultant X-ray luminosity (the sum of the background and the flare X-ray luminosities) is $L_{\rm X,ave}= 1.30 \times 10^{30}$~erg~s$^{-1}$ (the flare energy averaged over the period is $E_{\rm flare, ave}=1.84\times 10^{34}$~erg), which is in the range of typical values for classical TTSs.
The $r_{\rm crit}-E_{\rm ph}$ diagram for the time-averaged data is similar to the result of the flare with $E_{\rm flare}=10^{34}$~erg at the flare peak (orange line in Figure \ref{fig:rcrit}(a)).
We find that the time-averaged X-rays from flares can still have a prominent effect on disk ionization.




\begin{figure}[t]
    \centering
    \includegraphics[width=8cm]{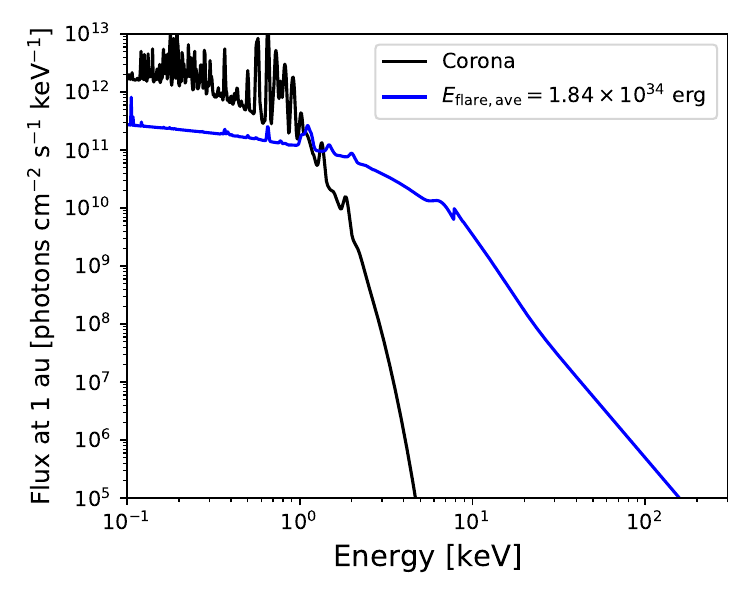}
    \caption{The X-ray spectra for the corona with $T=3\times 10^6$ K (black solid), and the 10-year averaged flare with $E_{\rm flare, ave}=1.84\times 10^{34}$ erg (blue).}\label{fig:specomp}
\end{figure}

Figure~\ref{fig:specomp} compares the time-averaged flare emission and the emission of a reference model of a steady stellar corona.
The time-averaged flare emission is constructed using $L_{X,\rm ave}$ and $E_{\rm flare,ave}$ as follows. 
We regard $E_{\rm flare,ave}$ as the representative flare energy for the given flare occurrence frequency and use it to calculate the flare temperature, $T_{\rm flare}$. The flare temperature determines the shape of the thermal spectrum $F_{\rm th}$ (at this point, the magnitude of $F_{\rm th}$ remains undetermined). To obtain the flare X-ray spectrum, we substitute $L_X = L_{X,\rm ave} - L_{\rm ch}$ in Equation~(\ref{eq_flux}) to find the nonthermal component for a given $f_{\rm nth}$. As a result, we get a time-averaged flare spectrum constrained by the time-averaged X-ray luminosity, which is shown as the blue line.
The steady stellar corona is assumed to have the temperature of approximately $3.0 \times 10^6$ K, which is in the range of typical values for classical TTSs like TW Hydrae \citep{Kastner2002, Gunther2007}. The emission from the steady coronal with the temperature and $L_{X}=L_{\rm ch}= 10^{30}$~erg~s$^{-1}$ is plotted by the black line.
The sum of the two components (black and blue lines) would represent the time-averaged X-ray spectrum of the flaring star.

Figure~\ref{fig:specomp} highlights the importance of the flare emissions in hard X-ray band from a star as a whole. The time-averaged flare produces a significantly stronger hard X-rays because it has a higher temperature than the steady corona and displays the nonthermal hard X-rays. We have confirmed that only the steady coronal X-ray emission does not affect the ionization rate at the disk midplane. Considering this result, we argue that a realistic modeling of the stellar X-ray emission is critical for better understanding the stellar impacts on the disk ionization state.

\section{Summary and Discussion} \label{sec:summary}

We constructed a model of X-ray spectra of stellar flares  based on the solar/stellar-flare theories and observations.
The rise and decay timescales of X-ray light curves, which determines the shapes of light curves, are described by functions of the flare energy $E_{\rm flare}$. This allows us to calculate the peak X-ray luminosity for a given flare energy in a way consistent with observations. The key factors that shape the X-ray spectra, which are the flare temperature and nonthermal emission, are also modeled. Our model enables us to better predict the hard ($\gtrsim 10$~keV) X-ray photon flux than before.

Using our X-ray model, we studied the response of a disk to flare X-rays by performing the radiative transfer calculations.
We generally found that flare X-rays affect the ionization rate down to $z=0.1R$ in a wide range of radii.
We also found that at the midplane, the hard ($\gtrsim 10$~keV) X-ray photons from flares with $E_{\rm flare} \geq 10^{36}$ erg have a significant contribution to ionization even at inner radii ($\sim$ a few au).
This is because they can penetrate into the denser gas without undergoing gas and dust absorption, even though they account for only a small fraction in the total X-ray flux.


In addition to the single-flare study, the time-averaged X-rays of multiple flares were found to certainly affect the disk ionization.
We generated the 10-year X-ray light curve using the flare occurrence frequency distribution based on solar and stellar observations.
Our 10-year-averaged result implies that X-ray flares should contribute to the disk ionization over long-time scales.

Our X-ray flare model contains some uncertain parameters, such as $f_X$ (Equation(\ref{eq_fx})) and $f_{\rm nth}$ (Equation (\ref{eq_frac})).
Theoretical models of stellar flare emissions, which incorporate the effect of nonthermal electrons, provide some hints to predict the possible range \citep[e.g.][]{Waterfall2019MNRAS,Waterfall2020MNRAS,Kimura2023ApJ}, but these parameters need to be constrained by observations. 
Simultaneous optical and X-ray observations are required to estimate the energy partition factor $f_X$. Radio observations which investigate the gyro-synchrotron emissions from nonthermal electrons may give some constraints on $f_{\rm nth}$.

This study greatly simplifies the radiative transfer of flare X-rays, which needs to be improved in future. For instance, we have ignored the presence of the accretion columns \citep[e.g.][]{Hartmann2016ARA&A} and disk winds \citep[e.g.][]{Pascucci2023ASPC} that can attenuate the stellar radiation. Recent star-disk interaction simulations by \citet{Takasao2022} find that the accretion columns and winds in the vicinity of the star can significantly block the starlight \citep[see also][]{Takasao2018}. 
A detailed modelling of the optical depth around the star is a key for model improvement.

Cosmic rays produced in stellar flares, which are ignored in this study, could play a role in the vicinity of the stars.
Recently, \cite{Brunn2023, Brunn2024} argued that energetic particles produced by flares can primarily ionize the inner disk at $<1$ au. Such a small scale structure is excluded in our simulations. To understand the stellar impacts in a wide range of radii, it will be important to reveal the roles of different ionizing sources including the flare-origin cosmic rays.
We note that there are similarities in modeling the flare X-rays between our and their studies, but the nonthermal X-ray emission is missing in their model.


Monitoring observations should facilitate the verification of X-ray driven chemistry in protoplanetary disks \citep{Henning2010, Waggoner2023}.
Currently, numerical codes that calculate the disk evolution with a large chemical network are available \citep[e.g.,][]{Furuya2014}. Implementing our X-ray flare model into such models will enable us to better compare theoretical predictions with observations.


H.W. is supported by JSPS KAKENHI grant No. JP24KJ0151.
S.T. is supported by the JSPS KAKENHI grant No. JP21H04487, JP22K14074, and JP22KK0043. 
K.F. is supported by the JSPS KAKENHI grant No. JP20H05847 and JP21H04487.

%




\appendix

\section{Effect of disk parameters}
\label{sec_app1}

To further understand the contribution of flare X-rays to disk ionization, we investigate the effects of disk mass and dust settling.
The change in disk mass, $M_{\rm d}$, affects the density structure of the disk, which control the X-ray attenuation by absorption.
The effect of dust growth is also important to have possible impacts on our results.
As dusts grow during disk evolution, they settle toward the midplane \citep[e.g.,][]{Weidenschilling1993}.
This effect is described with a dust settling parameter, $\epsilon$, which influences the X-ray opacity;
\begin{align}
 \sigma_X = \sigma_{\rm gas} + \epsilon f_{\rm b}(E) \sigma_{\rm dust},
\label{eq:dustopac}
\end{align}
where $f_{\rm b}$ is the self-blanket factor dependent on energy \citep{Bethell2011}.

\begin{figure}[!t]
    \centering
    \includegraphics[width=17cm]{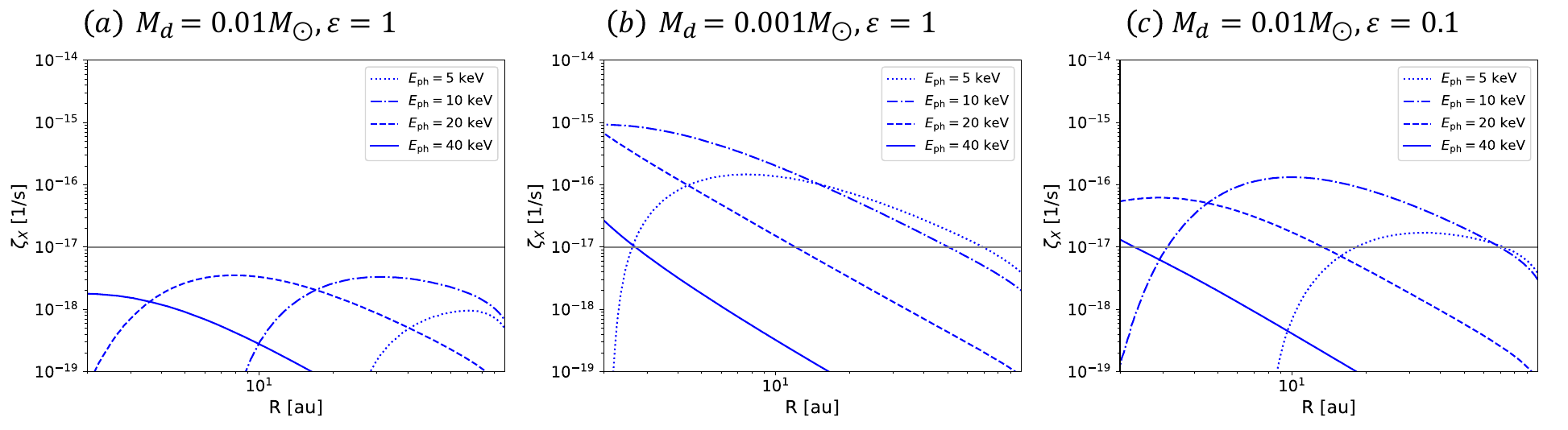}
    \caption{Comparison of the ionization rates at the midplane for the case of the flare peak with $E_{\rm flare}=10^{35}$ erg. (a) Our fidutial case with $M_d=0.01M_{\odot}$ and  $\epsilon=1$. (b) The case with a lower disk mass of $M_d=0.001M_{\odot}$. (c) The case with a lower settling parameter of $\epsilon=0.1$. }\label{fig:comp_diskparam}
\end{figure}

For the fiducial model (Section \ref{sec:d_model}), we set the disk mass with $M_{\rm d}=0.01 M _{\odot}$ and $\epsilon=1$ , where no dust settling occurs in the disk.
We first run the additional calculations with $M_{\rm d}=0.001 M_{\odot}$, keeping all other parameters unchanged.
Figure \ref{fig:comp_diskparam}(a) and (b) display the ionization rates at the midplane for our fiducial case with $M_{\rm d}=0.01 M _{\odot}$ and for the case with a lower disk mass of $M_{\rm d}=0.001 M _{\odot}$.
When $M_{\rm d}$ is lower, the X-ray photons experience less absorption within the disk.
As a result, they can easily penetrate down to the midplane to increase the ionization rates. 
In particular, we find that the X-ray photons with 5 and 10 keV can have an larger effect on ionization at inner radii compared to the fiducial case.
In this case, $\zeta_X > \zeta_{\rm CR,0}$ holds across all radii, so that the flare hard X-rays are the primary source of the ionization at the midplane.

Figure \ref{fig:comp_diskparam}(c) presents the ionization rate at the midplane for a lower dust settling parameter of $\epsilon=0.1$.
Since the lower $\epsilon$ significantly reduce the X-ray dust opacity, the ionization rate increases at the midplane, yielding $\zeta_X > \zeta_{\rm CR,0}$ at almost all radii.
These results show the significant impact of disk properties on the ionization states.

\section{Impact of X-ray source position}
\label{sec_app2}

\begin{figure}
    \centering
    \includegraphics[width=10cm]{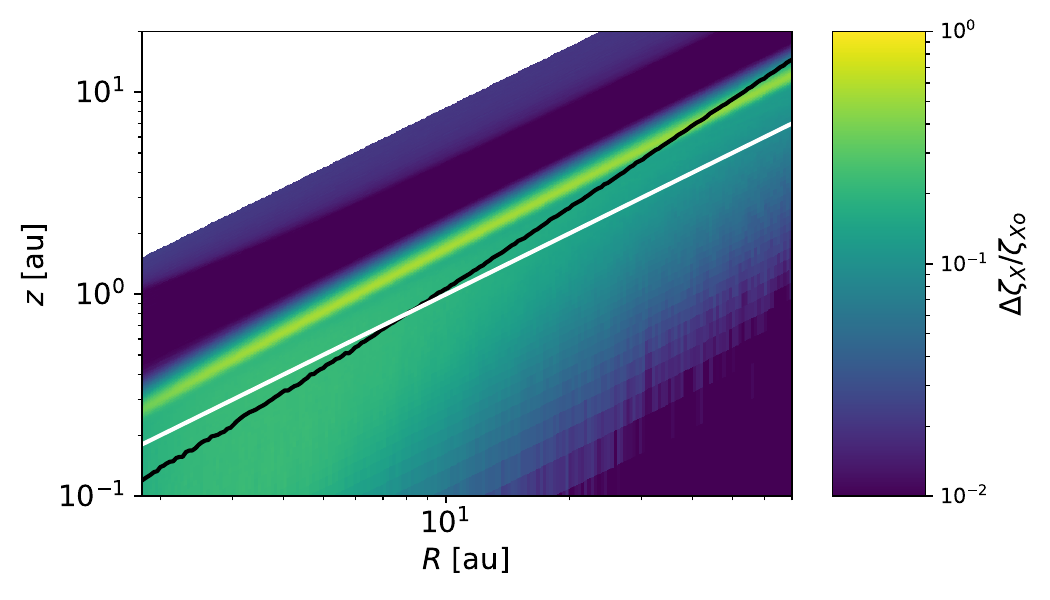}
    \caption{The impact of the position of an X-ray source. The color bar shows $\Delta \zeta_X / \zeta_{Xo}$. The result is for $E_{\rm ph}=20$ keV at the peak of flare with $E_{\rm flare}=10^{35}$ erg.}\label{fig:iratio_geo}
\end{figure}

One of the limitations of our model is the treatment of the flare geometry. 
In Section \ref{sec:disk} and Section \ref{sec:result}, we assume that the flare X-rays are uniformly emitted from a point source at $(R,z)=(0,0)$ for simplicity. 
However, in reality, flares have finite sizes and occur at specific locations on the stellar surface. 
We here discuss how the flare geometry affects the results.

Our model relates the flare spatial scale and the flare energy using the following relation:
\begin{align}
E_{\rm flare} \approx \frac{B^2}{8\pi}L^3 .
\label{eq_scale}
\end{align}
Assuming a flare region as a sphere with radius $R_{\rm f}$, namely $L^3 = 4\pi R_{\rm f}^3 /3$, we can express it as
\begin{align}
R_{\rm f} \approx 10^{11} \left(\frac{E_{\rm flare}}{10^{35} \mathrm{erg}}\right)^{1/3} \left(\frac{B}{50 \mathrm{G}}\right)^{-2/3} \hspace{2mm} \mathrm{cm}.
\end{align}
For $E_{\rm flare}=10^{35}$ erg, $R_{\rm f}$ is approximately equal to the stellar radius of $R_*=2.0 R_{\odot}$.
Considering this, we model the X-rays as being emitted from the center of mass of the flare rooted on the stellar surface.
We conduct simulations where X-rays from a flare with $E_{\rm flare} = 10^{35}$ erg are emitted from the point at $(R,z)=(0, 1.5 R_*)$. 
While previous studies adopt a source height of $z=10 R_{\odot}$ \citep[e.g.,][]{Igea1999}, Equation (\ref{eq_scale}) suggests that the flare size is expected to be smaller than this value.

To compare the ionization rates with the two cases for the X-ray source position, we define the relative change in the ionization rate as 
\begin{align}
\frac{\Delta \zeta_X}{\zeta_{Xo}} \equiv \frac{|\zeta_{Xo} - \zeta_{X1}|}{\zeta_{Xo}},
\label{eq:del_zetax}
\end{align}
where $\zeta_{Xo}$ and $\zeta_{X1}$ represent the ionization rates by the X-rays from the point $(R,z)=(0, 0)$ and $(R,z)=(0, 1.5 R_*)$, respectively. 
Figure \ref{fig:iratio_geo} shows $\Delta \zeta_X/\zeta_{Xo}$ for $E_{\rm ph}=20$ keV and $E_{\rm flare}=10^{35}$ erg.
When the X-rays are emitted from the point $(R,z)=(0, 1.5 R_*)$, the ionization rate at the scattering surface ($\tau_{\rm scat}=1$) is higher than that in the case of $(R,z)=(0, 0)$.
This increase is due to the steeper incidence angle on the scattering surface, which allows more X-rays to penetrate the disk. As a result, the ionization rate increases by a few tens of percent near the scattering surface and in the disk's inner region.
The result indicates that the ionization condition is not significantly affected by an X-ray source position.

\nocite{*}
\bibliography{washinoue+2024}


\bibliographystyle{aasjournal}



\end{document}